\documentclass{article}

\usepackage{graphicx} % Required for inserting images
\usepackage{authblk}
 
\makeatletter
\renewcommand{\AB@affilsepx}{, \space} % default is \\ \Affilfont
\makeatother
\usepackage[hidelinks]{hyperref}
\usepackage{amsmath}
\usepackage{float}
\usepackage{xcolor}
\usepackage{subcaption}
\usepackage{cleveref}
\usepackage{rotating}
\usepackage{float}
\usepackage{lineno}

\hypersetup{
    colorlinks,
    linkcolor={red!50!black},
    citecolor={blue!50!black},
    urlcolor={blue!80!black}
}

\usepackage[top=1.3cm, bottom=2.0cm, outer=2.4cm, inner=2.4cm, heightrounded,
marginparwidth=1.5cm, marginparsep=0.4cm, margin=2.5cm]{geometry}

\title{\textbf{Machine Learning Power Week 2023: Clustering in Hadronic Calorimeters}}

\author[7]{M. Al Halabi}
\author[9]{M. Bajdel}
\author[8,9]{J.P. Bormans} % j.bormans@gsi.de
\author[1, 2]{H. Bossi (ed)}
\author[7]{M. Calmon Behling}
\author[8]{F. Ehmann}
\author[5, 6]{N. G\"otz} % goetz@itp.uni-frankfurt.de
\author[7]{J. Jung}
\author[11, 12]{R. Kavak}
\author[7]{M. Kr\"uger} % mkrueger@ikf.uni-frankfurt.de
\author[6, 15]{R. Lakos} % lakos@fias.uni-frankfurt.de
\author[13]{A. Lauterbach}
\author[2]{P. Mccormack (ed)}
\author[6]{A. Mithran}
\author[3,16]{D. Murnane (ed)}
\author[10]{M. Nolte} % mathis.nolte@uni-jena.de
\author[7]{T.S.~Rogoschinski} % rogoschinski@ikf.uni-frankfurt.de
\author[7]{J. Scharf} % jan.scharf@stud.uni-frankfurt.de
\author[6]{O. Tyagi} % otyagi@fias.uni-frankfurt.de
\author[4]{L. Våge (ed)}
\author[10,14]{M. Valialshchikov}
\author[13]{S. Wagner}

\affil[1]{Yale University, New Haven, CT, USA}
\affil[2]{Massachusetts Institute of Technology, Boston, MA
, USA}
\affil[3]{Niels Bohr Institute, Denmark}
\affil[4]{Imperial College, London, UK}
\affil[5]{Institute for Theoretical Physics, Goethe University Frankfurt, Department of Physics, Frankfurt, Germany}
\affil[6]{Frankfurt Institute for Advanced Studies, Frankfurt am Main, Germany}
\affil[7]{Institut für Kernphysik, Johann Wolfgang Goethe-Universität Frankfurt, Frankfurt, Germany}
\affil[8]{Institute for Nuclear Physics, Dept. of Physics, Technische Universität Darmstadt, Germany}
\affil[9]{GSI Helmholtzzentrum f\"ur Schwerionenforschung, Darmstadt, Germany}
\affil[10]{Institut für Optik und Quantenelektronik, Friedrich-Schiller Universität, Jena, Germany}
\affil[11]{Physikalisches Institut, Universität Heidelberg, Heidelberg, Germany}
\affil[12]{Institut für Theoretische Physik, Universität Heidelberg, Heidelberg, Germany}
\affil[13]{Institut für Angewandte Physik (IAP), Johann Wolfgang Goethe-Universität Frankfurt, Frankfurt, Germany}
\affil[14]{Helmholtz Institute Jena, Jena, Germany}
\affil[15]{Institute of Computer Science, J. W. Goethe University, Frankfurt (Main), Germany}
\affil[16]{Lawrence Berkeley National Lab, Berkeley, CA,  USA}

\begin{document}
% \linenumbers
\maketitle

% write the abstract here 
\begin{center}
    \textbf{Abstract}\\
    In both high-energy physics and industry applications, a crowd-sourced approach to difficult problems is becoming increasingly common. These innovative approaches are ideal for the development of future facilities where the simulations can be publicly distributed, such as the Electron-Ion Collider (EIC). In this paper, we discuss a so-called ``Power Week" where graduate students were able to learn about machine learning while also contributing to an unsolved problem at a future facility. Here, the problem of interest was the clustering of the forward hadronic calorimeter in the foreseen electron-proton/ion collider experiment (ePIC) detector at the EIC. The different possible approaches, developed over the course of a single week, and their performance are detailed and summarised. Feedback on the format of the week and recommendations for future similar programs are provided in the hopes to inspire future learning opportunities for students that also serve as a crowd-sourced approaches to unsolved problems. 
\end{center}

% \newpage
\section{Introduction}
In recent years, both private companies and scientific collaborations have turned to crowd-sourcing platforms in order to implement real solutions to problems that are challenging both computationally and scientifically ~\cite{Kasieczka:2021xcg,icecube-neutrinos-in-deep-ice,Krause:2024avx,Bhimji:2024bcd,Aarrestad:2021oeb}. Such approaches have proven successful at brainstorming new, innovative solutions to long-standing problems. One barrier to such crowd-sourcing being widely adopted in high-energy physics (HEP) contexts is the difficulty of releasing data publicly. Future facilities where details of detector configurations, technologies, resolution, etc., are not yet finalised, offer one potential context within HEP that avoids this difficulty. In addition, crowd-sourcing such approaches allows for innovative and creative solutions to long-standing problems. \\

The Electron-Ion Collider (EIC) is a planned particle accelerator facility that will collide electrons of various energies with protons and nuclei~\cite{Accardi:2012qut,AbdulKhalek:2022hcn}. The facility will be constructed at Brookhaven National Laboratory (BNL) on Long Island in the United States and will make use of the existing ion ring from the Relativistic Heavy Ion Collider (RHIC) facility. There is one full-acceptance general-purpose detector, referred to as the electron/Proton Ion Collider Experiment (ePIC), currently planned for the EIC. Some physics goals of the EIC include; determining how the properties of nucleons emerge from partons and their interactions, how partons are distributed within nucleons, how hadrons are formed from the confinement of partons, and more~\cite{AbdulKhalek:2022hcn}. In order to accomplish these physics goals, the EIC will be constructed as one of the most sophisticated facilities to date. Towards this aim, there are large efforts in place in order to incorporate advanced techniques in the design, operation, reconstruction simulation, and physics analysis at the EIC~\cite{Allaire:2023fgp}. \\

As a part of these efforts, an ML Power Week was organised as a part of the Helmholtz Graduate School for Hadron and Ion Research, which took place from July 18th - 22nd, 2023, in Schmitten, Germany. This ``Power Week" focused on the development of a clustering algorithm for the ePIC forward hadronic calorimeter. The Power Week was organised as a private Kaggle Competition~\cite{hgs-hire-power-week-an-epic-competition} where participants would compete in groups in order to create the clustering algorithm with the best performance (see Section \ref{sec:perfMetric} for more details). In this paper, the setup and results of this competition are summarised. In Section \ref{sec:dataset}, the data set and setup of the challenge will be described. In Section \ref{sec:previous}, previous approaches to clustering-based problems will be presented. In Section \ref{sec:Approaches}, the various approaches that were attempted to solve this problem will be discussed. In Section \ref{sec:discussion} the results will be discussed in context and some recommendations for the format of future workshops will be provided. Finally, in Section \ref{sec:summary}, the Power Week and its main takeaways will be summarised. 

\section{Dataset and Challenge}\label{sec:dataset}
For this Power Week, a simulation of a so-called ``particle gun" hitting the ePIC forward hadronic calorimeter was used. In the following sections, the details of the calorimeter (Section \ref{subsec::calo}), the simulation used (Section \ref{subsec:sim}), as well as the setup of the competition (Section \ref{subsec:comp}) and the performance metric used (Section \ref{sec:perfMetric})  will be discussed.
\subsection{The ePIC Forward Hadronic Calorimeter}\label{subsec::calo}
The forward electromagnetic and hadronic calorimeters in the hadron-going endcap of the ePIC detector are essential subdetectors for accomplishing its physics goals. This is primarily due to the fact that the incoming proton/nuclei will have significantly higher kinetic energy than the electron, and therefore, a majority of hadrons will occur in the forward region. The measurement of particles in the forward region will additionally rely on the calorimeter system as tracking momentum and resolution worsens at very forward values of the pseudorapidity. 

\begin{figure}[ht!]
    \centering
    \includegraphics[width = 0.6\textwidth]{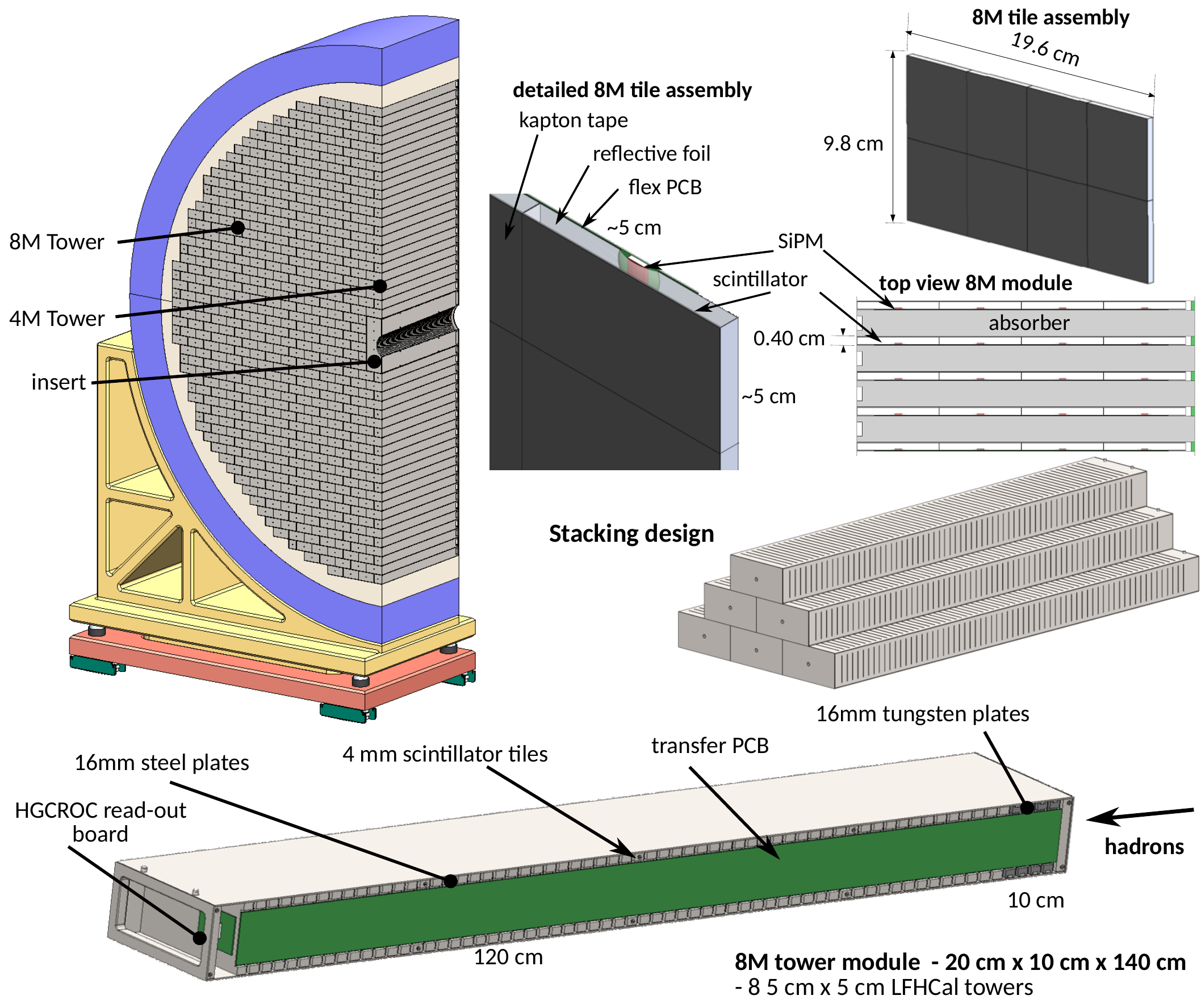}
    \caption{3D rendering of the design of the LFHCal (left) and a zoomed-in rendering of a single 8M module (bottom). Individual components of the 8M module are shown in the top right corner, as well as the stacking concept (middle right).  Please note that this is the design as of R\&D proposal \# 107, submitted July 2023, and as a result and has since changed. See \href{https://wiki.bnl.gov/EPIC/index.php?title=Forward_Hcal}{https://wiki.bnl.gov/EPIC/index.php?title=Forward\_Hcal} for the most up to date design. Note that both of these renderings originally appeared in the  R\&D proposal \# 107.}
    \label{fig:lfhcalSpecs}
\end{figure}

In this paper, the design of the forward hadronic calorimeter of the ePIC detector will refer to the design as it existed as of the R\&D proposal \# 107~\footnote{See \href{https://wiki.bnl.gov/conferences/images/8/8f/ERD_107_2023_submitted.pdf}{https://wiki.bnl.gov/conferences/images/8/8f/ERD\_107\_2023\_submitted.pdf} for details.} that was submitted in July 2023. These conditions are subject to change as detector conditions evolve over time in conjunction with developments in the other subdetectors of ePIC. The detector part of interest is commonly referred to as the longitudinally segmented forward hadronic calorimeter (LFHCal) and is a steel-scintillating sampling calorimeter with its first segment made out of tungsten and scintillators, serving as a collimator. The LFHCal consists of two movable half-discs such that the calorimeter can be wheeled out of the way to give access to the internal detector components. This disc, one half of which is shown in the left-hand side of Figure \ref{fig:lfhcalSpecs}, has a radius of 270 cm, an active depth of 130 cm, and 8916 towers with a 5 cm X 5 cm front face. Each tower consists of so-called 8M modules arranged in a Lego-like structure. One such module is shown in the right-hand side of Figure \ref{fig:lfhcalSpecs}, and comprises 8 towers in a 4 by 2 tower arrangement, as seen in the top left of  Figure \ref{fig:lfhcalSpecs} displaying a single 8M module layer and its segmentation. Each tower  consists of 7 different readout channels - forming 7 independent layers in the z-direction. The asymmetric structures immediately surrounding the beam pipe have been added in order to increase the coverage of the calorimeter. See the eRD \# 107 linked for more details.\footnote{For the updated LFHCal design see \href{https://wiki.bnl.gov/EPIC/index.php?title=Forward_Hcal}{https://wiki.bnl.gov/EPIC/index.php?title=Forward\_Hcal}.}

\subsection{The Simulations}\label{subsec:sim}
\begin{figure}[ht!]
    \centering
    \includegraphics[width = 0.49\textwidth]{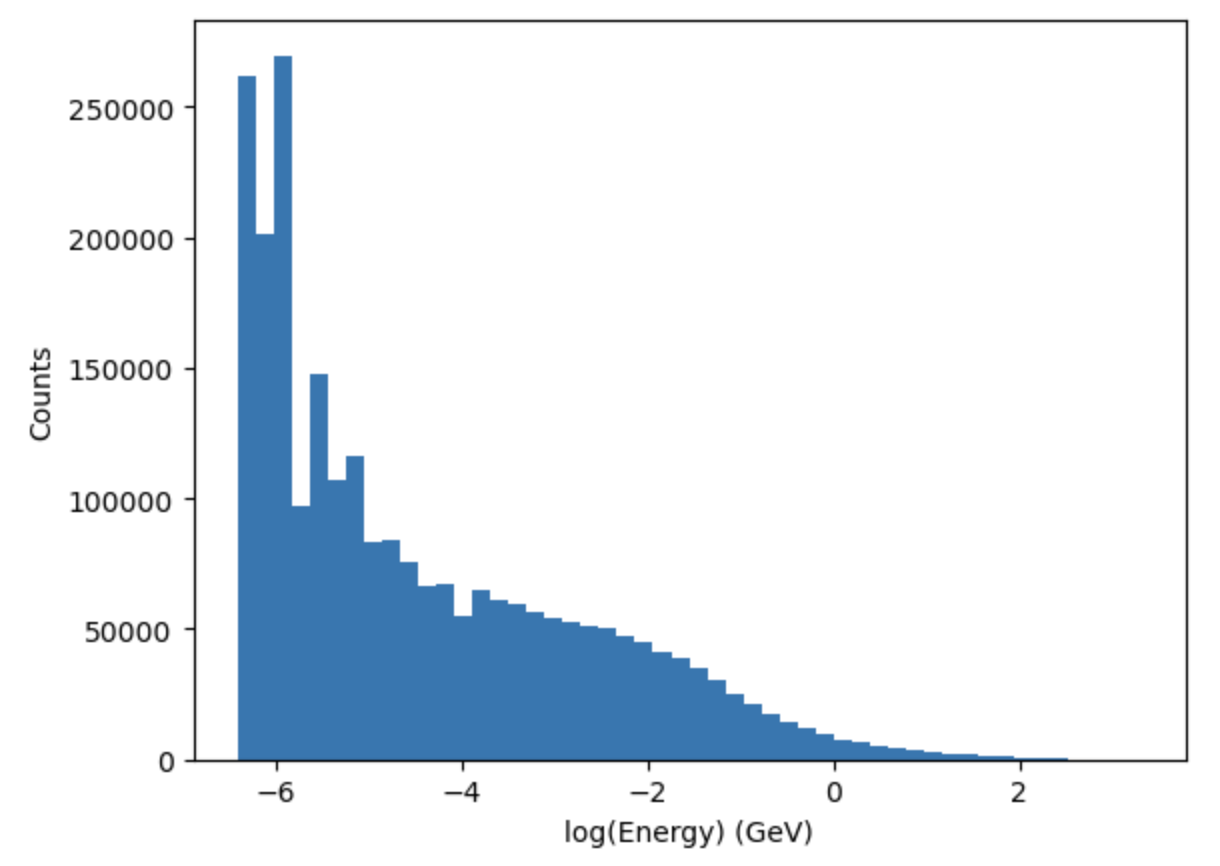}
    \includegraphics[width = 0.49\textwidth]{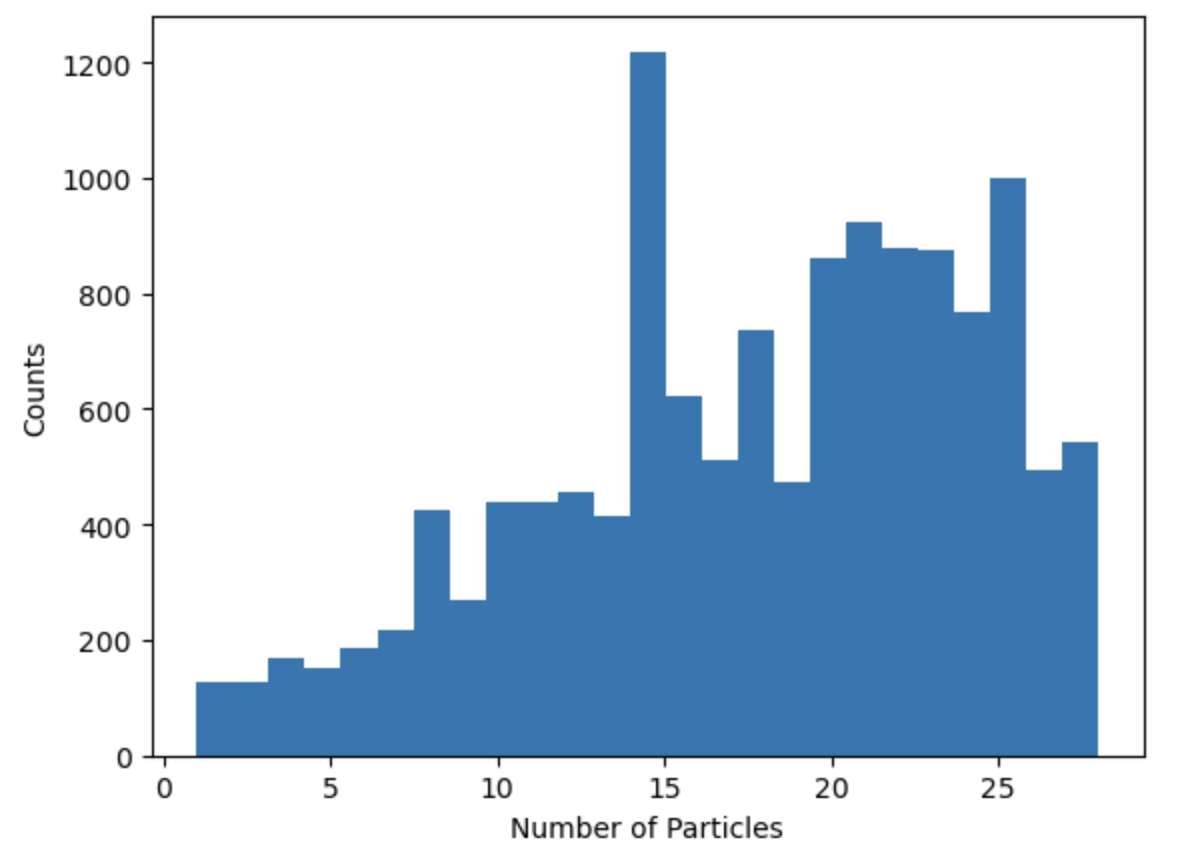}
    \caption{Distribution of the logarithm of the energy distribution for all hits in the training sample (left) as well as the distribution of the number of particles per event in the training distribution (right).}
    \label{fig:1DDistributions}
\end{figure}

The simulations for this Power Week were provided by the ePIC collaboration and reflect the geometry of the LFHCal as described above. In these simulations, 100,000 events were simulated using the merged ``particle gun" style with neutrons, protons, and pions fired from the interaction point that is 3.58 m from the surface of the LFHCal~\footnote{These 100,000 events were then split into a training and test sample for the competition, which is described in Section \ref{subsec:comp}.}. Note that these simulations were completed \textbf{without} taking into account particle interactions with the electromagnetic calorimeter that is located immediately in front of the LFHCal. The particle multiplicities and energies in these events are set to reflect semi-realistic conditions for ePIC data-taking, averaging around 20 particles per event in the training sample, as can be seen in the right panel of Figure \ref{fig:1DDistributions}. In the left panel of Figure \ref{fig:1DDistributions} is the energy distribution for all hits in the training sample. 

\begin{figure}[ht!]
    \centering
    \includegraphics[width = 0.49\textwidth]{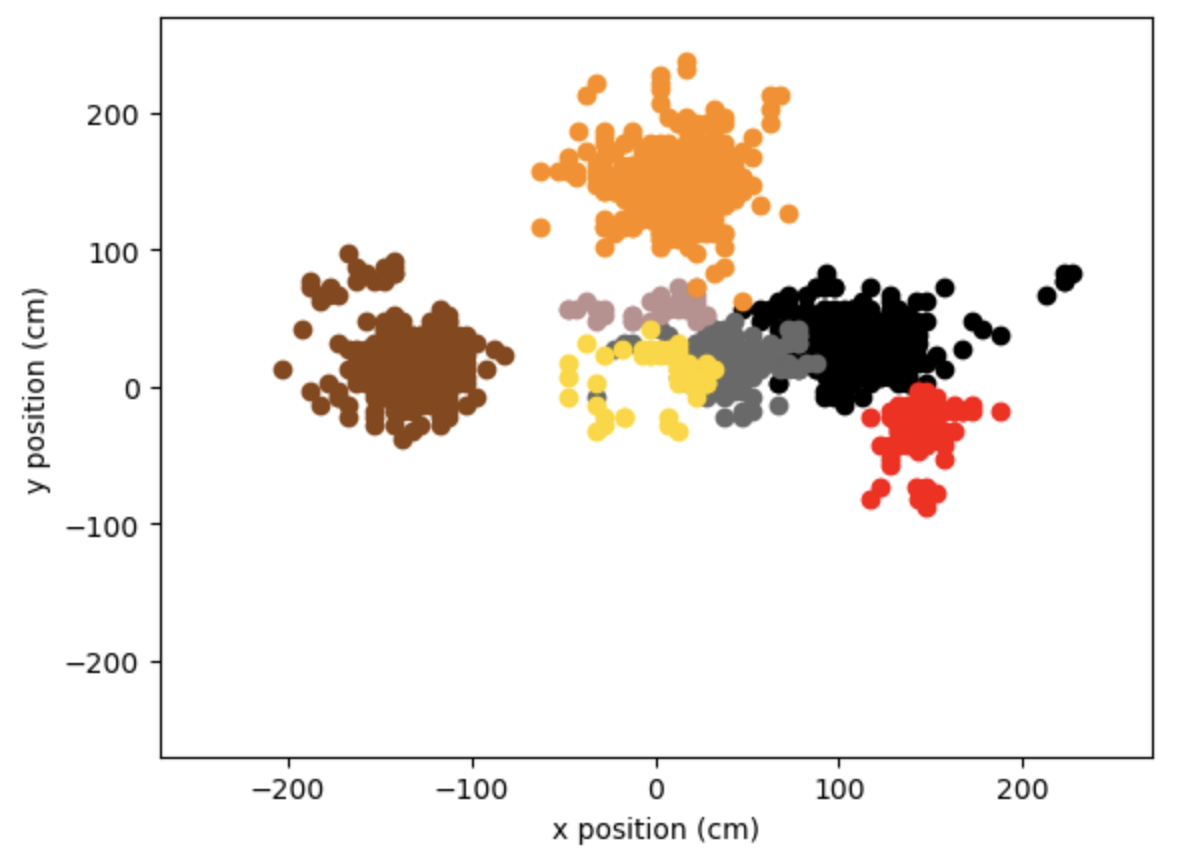}
    \includegraphics[width = 0.49\textwidth]{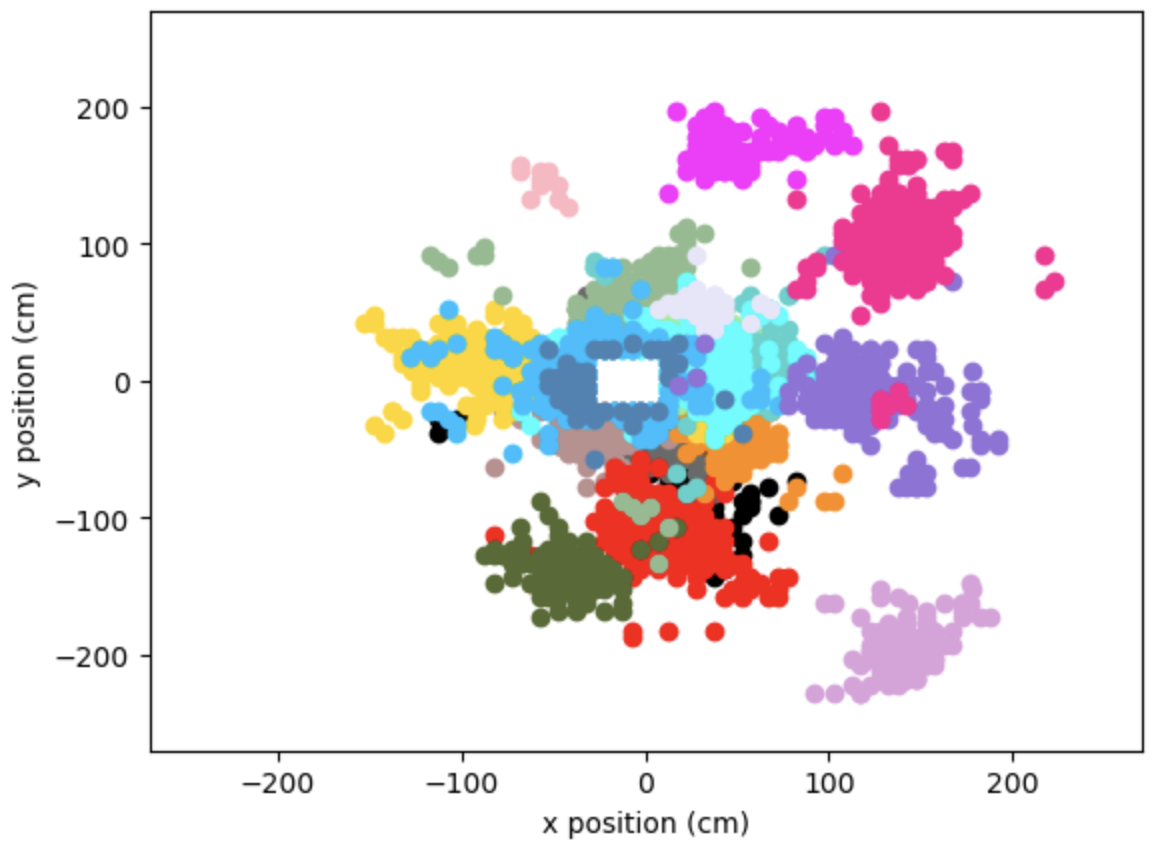}
    \caption{Distribution of hits for various clusters in the $xy$ plane for an event with low particle multiplicity (left) and high particle multiplicity (right). Note that each unique colour marks a different true cluster.}
    \label{fig:typicalEvent}
\end{figure}

The relative difficulty of the clustering problem varies largely from event to event. For example, the view  of clusters in the  $xy$ plane for two different  events with drastically different multiplicities is shown in Figure \ref{fig:typicalEvent}. From these two sample events, a few observations can be made. Firstly, even by eye, the difficulty of the clustering problem drastically increases with multiplicity. Secondly, the average event contains many hits very close to the beam pipe. Thirdly, even in events  with low multiplicity, there is a geometric overlap of clusters in the $xy$ plane. Note that this problem would become even more difficult in a more realistic simulation where particles may shower in the forward electromagnetic calorimeter before reaching the LFHCal. For the purposes of development and simplicity, the clustering problem, in this case, is considered only in the LFHCal with the expectation that this will be extended to include the other calorimeters at a later date.

\subsection{Competition Setup}\label{subsec:comp}
The competition was set up using a Kaggle format~\cite{hgs-hire-power-week-an-epic-competition} where participants must have an invite link in order to access the data and participate. The computing resources available on Kaggle, including GPU and CPU resources, were available to the participants. The competition was opened on July 18th, 2023, and closed on July 21st, 2023, at 23:59. Seven teams participated in the competition, with an average of 3 participants per team for a total of 21 participants. Participants were given access to a training data sample of 50,000 events with the ``true" cluster ID provided. In addition, participants were provided with a test sample of 10,000 events without true cluster ID labels in order to test and evaluate their approaches. Note that the sample provided from the ePIC collaboration was not used in its entirety in order to reduce the time needed to train and test algorithms. Teams were additionally limited to 20 submissions per day.  As is common on Kaggle, the participants could only view their performance on part of the test sample to avoid overfitting. Their performance on the full test sample was released at the end of the competition. The competition received 200 submissions over the course of the week. A benchmark algorithm was used with K-means clustering as implemented in \texttt{scikit-learn}~\cite{scikit-learn} with 40 clusters. For the performance of this benchmark, refer to Table \ref{tab:perfComp}. Participants were additionally given a number of code examples in order to help inspect events, evaluate performance, etc~\footnote{See \href{https://github.com/wpmccormack/ePIC_Clustering_2023}{https://github.com/wpmccormack/ePIC\_Clustering\_2023} for more details.}. Participants were also provided with a simplified approach to the clustering algorithm based on Section 3.4 of Ref~\cite{CMS:2017yfk} in order to perform the clustering in their hadronic calorimeters, which will be reviewed in Section \ref{sec:previous}. 

\subsection{Performance Metric}\label{sec:perfMetric}
There are many possible metrics that could be used in order to evaluate the performance of a calorimeter clustering algorithm. Two possible metrics are the \textit{homogeneity}, which measures the extent to which each cluster contains hits from a single particle, and the \textit{completeness}, which measures the extent to which all hits from a given particle are captured in a single cluster. These are defined in \cite{utt2014fuzzy,rosenberg2007v}, as functions of the conditional entropy $H(C|G)$ of a clustering assignment $C$ given a class truth label $G$, or a class truth label $G$ given a clustering assignment $C$, respectively. Both are then normalised by the joint entropy $H(C,G)$.

\begin{equation}\label{eq:homogeneity}
     hom = 1 - H(C|G)/H(C,G)
\end{equation}

\begin{equation}\label{eq:completness}
    com = 1 - H(G|C)/H(C,G)
\end{equation}

The joint and conditional entropies are calculated simply by counting the occurrences of hits $p_i$ in each cluster or class $i$, and the common occurrences of hits $p_{ij}$ in cluster $i$ \textit{and} class $j$:

\begin{align}
    H(C|G) = -\sum_{i\in C, j\in G} p_{ij} \log\frac{p_{ij}}{p_j} , && H(G|C) = -\sum_{i\in C, j\in G} p_{ij} \log\frac{p_{ij}}{p_i} , && H(C,G) = -\sum_{i\in C, j\in G} p_{ij} \log p_{ij}
\end{align}

\noindent One may also consider other metrics such as code run time, physical intuition, and performance in a given energy range or particle species as additional ways to measure the performance. However, for the purposes of comparison between models and for the Kaggle competition, a single number must be chosen in order to evaluate the models. For this purpose, a so-called \textit{validity score} was used. Proposed in \cite{rosenberg2007v}, this validity score ($V_{score}$) is the harmonic mean of the homogeneity ($H$) and the completeness ($C$), 

\begin{equation}\label{eq:vscore}
  V_{score} = 2\cdot \frac{hom \cdot com}{hom + com}  
\end{equation}

\noindent This is analogous to the F-score commonly used in binary classification tasks. This metric was chosen as it requires that a single algorithm must have good homogeneity and completeness in order to perform well - that is an homogeneity or completeness of 0 gives a V-score of 0, with a maximum possible validity score of 1. To better capture the physics case of clustering in the hadronic calorimeter, the standard V-score was amended to include a measure of the importance of each clustered hit, leading to a \textit{weighted $V_{score}$}. We choose to weight hits linearly in the energy deposited in the hit, such that incorrectly clustering low-energy hits insubstantially affects the weighted $V_{score}$, compared with clustering high-energy hits from different particles or splitting the hits in a single high-energy particle shower. For the implementation of this weighted validity score in code, refer to the \href{https://github.com/wpmccormack/ePIC_Clustering_2023/blob/master/epic_clustering/scoring/scoring_functions.py}{github repository}. An effort to include this weighted version of the $V_{score}$ natively within \texttt{scikit-learn}~\cite{scikit-learn} is ongoing, and will make inclusion of this score in future challenges straightforward.

% previous approaches
\section{Previous Approaches}\label{sec:previous}
In this section, previous ``traditional approaches" to calorimeter clustering will be discussed. In any machine learning investigation, it is important to first build the domain knowledge necessary to complete the task. In order to provide this background, a selection of these previous approaches were presented in a series of lectures in the beginning of the Power Week. Whether well established or recently developed, these approaches provided inspiration and a physical intuition for the explored approaches that will be discussed in the next section. To limit the scope of this discussion, we will focus on clustering algorithms employed at the Large Hadron Collider (LHC), though many other experiments - for example those at the RHIC - also of course employ clustering techniques. 

Calorimeter clustering with the goal of grouping all of the energy deposits of a single particle together is often a component of a broader particle reconstruction scheme.
Both the CMS~\cite{CMS:2008xjf} and ATLAS~\cite{The_ATLAS_Collaboration_2008} experiments deploy ``particle flow'' algorithms with the goal of reconstructing every particle created in considered LHC events~\cite{CMS_PF, ATLAS_PF}.
Particle flow algorithms utilize different components of detectors to distinguish particles and measure their properties.
For example, electrons can be identified by combining a charged particle track and a cluster in an electromagnetic calorimeter, while charged hadrons can be reconstructed by combining a charged particle track with a cluster in a hadronic calorimeter.
The use of a particle flow algorithm can improve resolution for objects such as jets, especially relative to applications of jet algorithms to calorimeter cells alone.
This is possible because different detector components have superior resolution in different energy regimes; track momentum measurements are more accurate than calorimeter measurements for lower energy particles, and the opposite is true for higher energy particles.

\begin{figure}
    \centering
    \includegraphics[width=\linewidth]{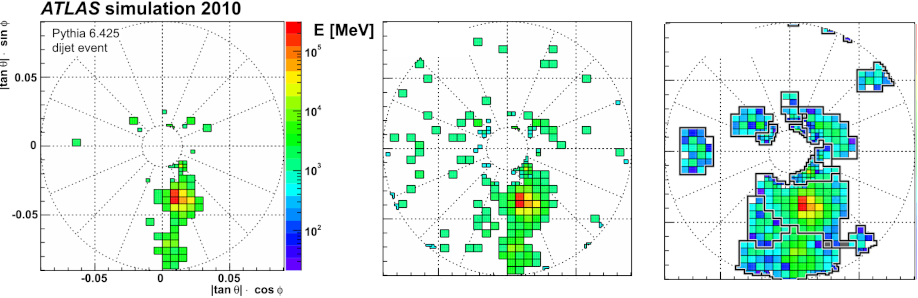}
    \caption{Example of clustering dijets in the ATLAS detector. Highly energetic cells act as seeds for clusters, as shown to the left. The clusters are grown by first adding the energetic, and then the less energetic neighbours. Figure adapted from \cite{clusteringFigure}.}
    \label{fig:clusteringExample}
\end{figure}

The clustering algorithms of ATLAS, CMS and ALICE all share the same main strategy for clustering in their calorimeter(s). They find cluster seeds by identifying the cells where the energy deposited is above a certain threshold. The clusters are grown by adding neighbouring cells in the same or neighbouring detector layers to the clusters. An illustration of this is shown in \autoref{fig:clusteringExample}. If one cluster contains several seeds, it can be split into several clusters. The details of these steps vary for each experiment, as they have slightly different detector geometries and experimental priorities. Details of the clustering algorithms of ATLAS, CMS and ALICE can be found in \cite{Lampl:2008zz}, \cite{cmsClustering}, and \cite{ALICE:2022qhn}, respectively. LHCb uses a largely similar approach, with the exception that the information is stored in a graph structure. Hits are nodes, and connections between hits are expressed in terms of edges. The largest draw of a graph structure is the increased computational efficiency that has been demonstrated \cite{Canudas:2022lsf}.

\par In recent years, the application of machine learning algorithms to the problem of calorimeter clustering has become increasingly more popular. For example, a graph neural network (GNN) was applied for end-to-end reconstruction in the CMS High-Granularity Calorimeter (HGCAL), which will replace the endcaps of the current CMS calorimeter system~\cite{CMS_HGCAL}. This GNN-based approach used the ``Object Condensation'' loss function~\cite{Kieseler_2020} and ``GravNet'' graph convolutional layer. When applied to a dataset consisting of the decay products of two $\tau$~leptons impinging on the HGCAL, the model achieved high-performance instance segmentation~\cite{Bhattacharya_2023}.

\section{Explored Approaches}\label{sec:Approaches}
In the following subsections, each of the broad categories of approaches explored during the Power Week as well as their advantages and disadvantages will be discussed. For the purposes of comparison the speed and performance (as defined in Section \ref{sec:perfMetric}) will also be discussed. For clarity, the final solution submitted from each team will be assigned a letter $A$-$G$ and will be referred to as ``submission $A$", for example. When referring to the group members themselves we will use ``Group $A$", for example.

\subsection{Non-ML Methods}
One class of explored approaches during the Power Week are non-ML methods. These methods have the principle benefit of being transparent and physics-motivated by design. Additionally, due to the additional time required to train ML methods, these methods may often be faster to implement and alter. But when comparing ML and non-ML algorithms, inference time is usually even more important, especially given the high data-streaming rates and low latency requirements of modern facilities. Two Non-ML methods were explored by workshop participants. The first is a CMS-inspired approach, which builds off of the algorithm described in Section \ref{sec:previous}. The second is an application of the jet clustering algorithm anti-$k_{\rm T}$ (Section \ref{subsubssec_antikt}) to the clustering problem.
The CMS algorithm achieved a score of 0.95499, making it the second to lowest scoring solution, as shown in \autoref{tab:perfComp}. As the algorithm is fairly sensitive to certain parameters and it was not optimised, the results are not compared further, but the algorithm can be found in \cite{epicGitRepo}. An overview of the submitted algorithms and their performance can be found in \autoref{tab:kmeans_algo_comparison}. 

\subsubsection{Anti-\texorpdfstring{$k_{\rm T}$}{kT}}\label{subsubssec_antikt}
Jet clustering algorithms are widely used in high-energy physics to group particles into a jet, but can also be used to cluster energy hits within a calorimeter. For example, recent studies with the electromagnetic pixel calorimeter prototype EPICAL-2 \cite{c_EPICAL_paper} deploy a jet-finding algorithm to cluster pixel hits produced by an electromagnetic shower. A similar approach was explored by Group \textit {G} for the case of hadronic showers produced in the LFHCal. Here, the anti-$k_{\mathrm{T}}$ jet-finding algorithm as implemented in the \texttt{fastjet} library \cite{Cacciari:2008gp,Cacciari_2012}  was applied in order to cluster the hits in the LFHCal. The anti-$k_{\mathrm{T}}$ algorithm is a sequential recombination algorithm that prioritizes the clustering of high $p_{\rm T}$ particles within a given resolution (or radius) parameter $R$. Typically, the anti-$k_{\mathrm{T}}$ algorithm clusters particles by sequentially adding particles to so-called pesudo jets via their associated four-vector ($p_{x}^{i},  p_{y}^{i},  p_{z}^{i} ,  E^{i}$). In this implementation, however, hits in the LFHCal are assigned to a pseudo jet $i$ according to the corresponding Cartesian position ${(x,y,z)}$ and the energy $E$.

% ================================================== 
% 
% ==================================================
\begin{figure}[t]
  \begin{center}
    \begin{minipage}[t]{0.49\textwidth}
      \begin{center}
        \includegraphics[width=\textwidth]{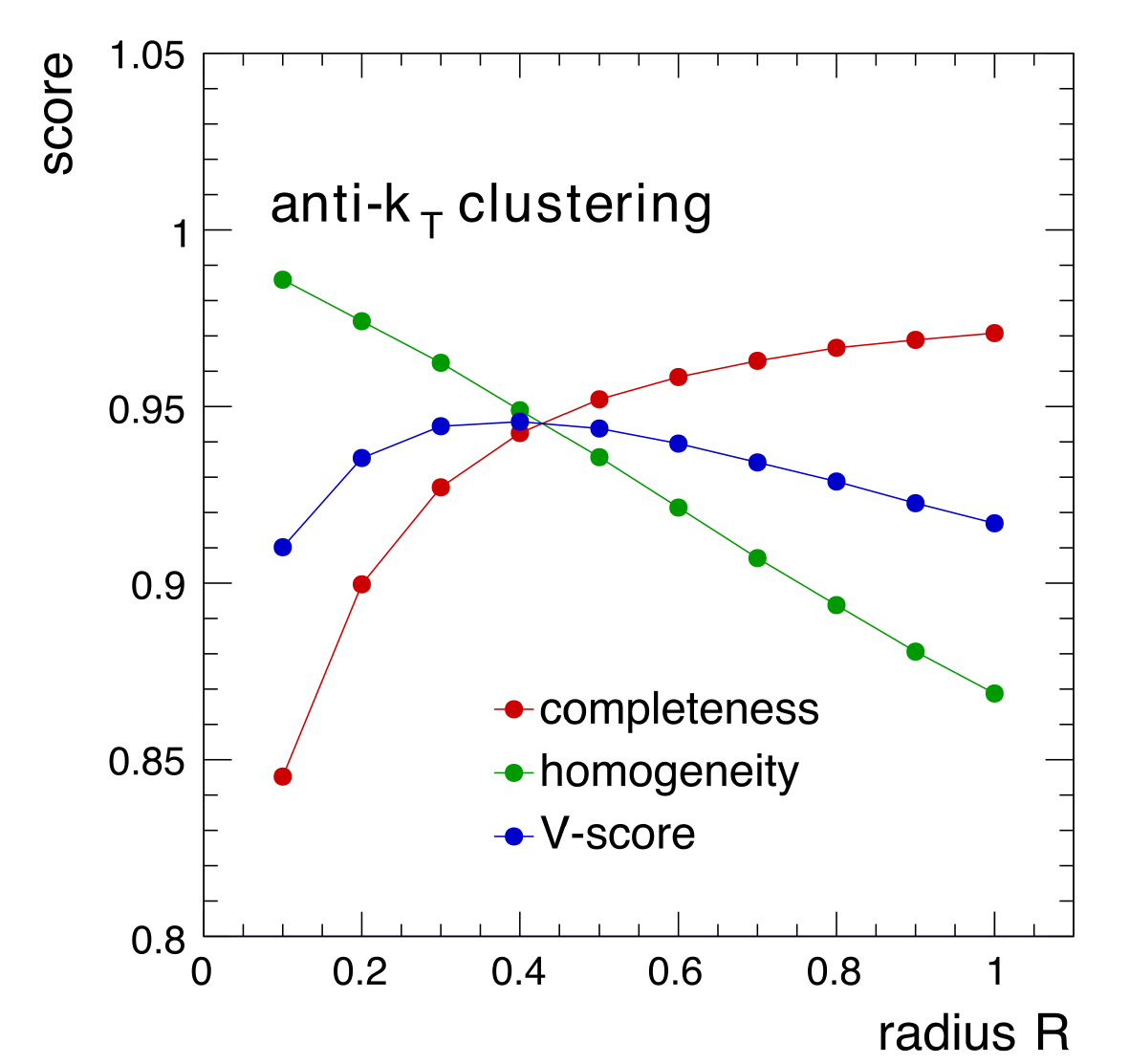}
      \end{center}
    \end{minipage}
    \begin{minipage}[t]{0.49\textwidth}
      \begin{center}
        \includegraphics[width=\textwidth]{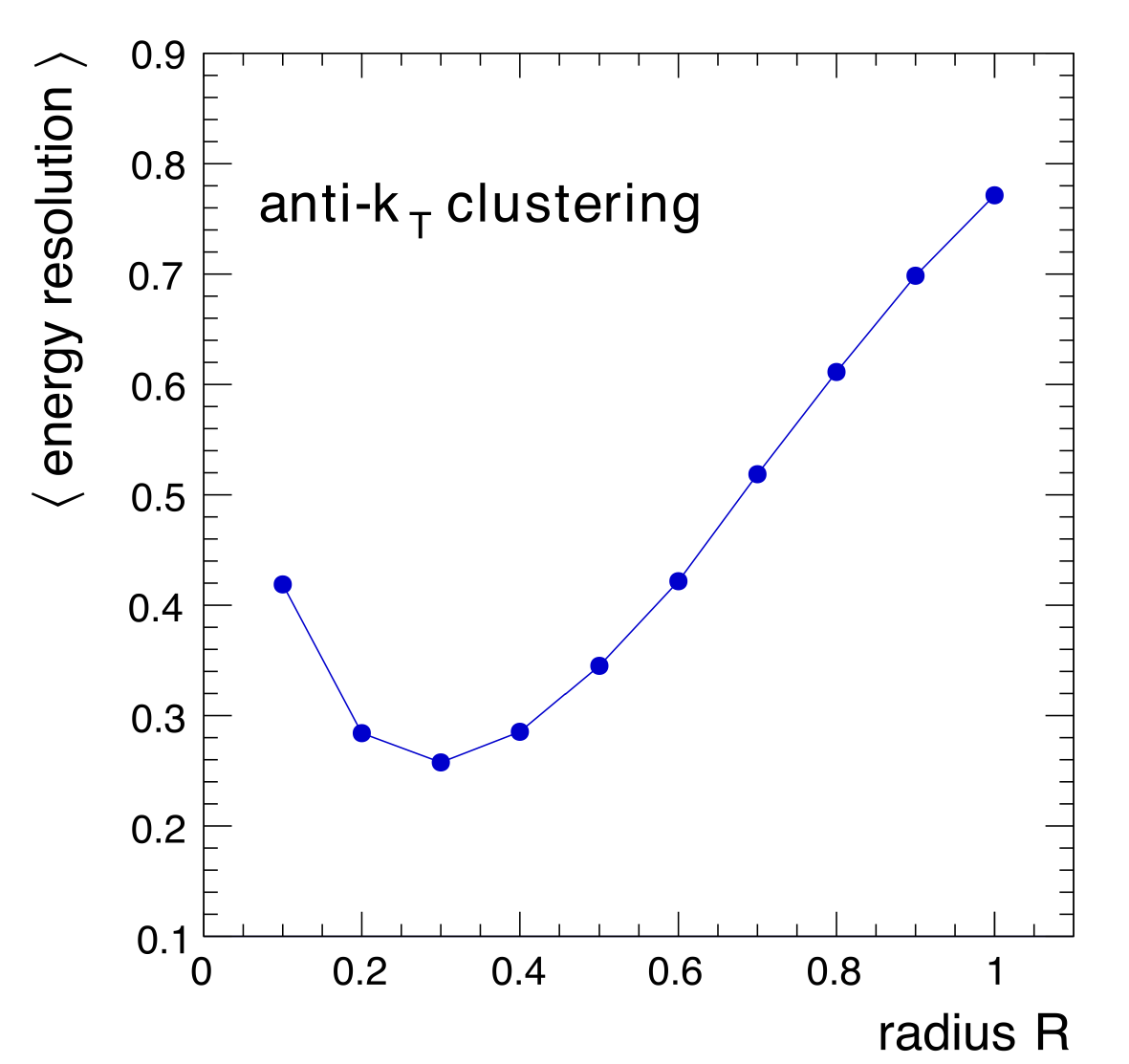}
      \end{center}
    \end{minipage}
    \caption{Performance metrics (left) and relative energy resolution (right) as a function of 'jet' radius $R$. Figure produced by Group \textit{G}.}
    \label{fig:kT_algorithm_vScore_and_egy_reso_final}
  \end{center}
\end{figure}

\begin{figure}[t]
  \begin{center}
        \includegraphics[width=1.0\textwidth]{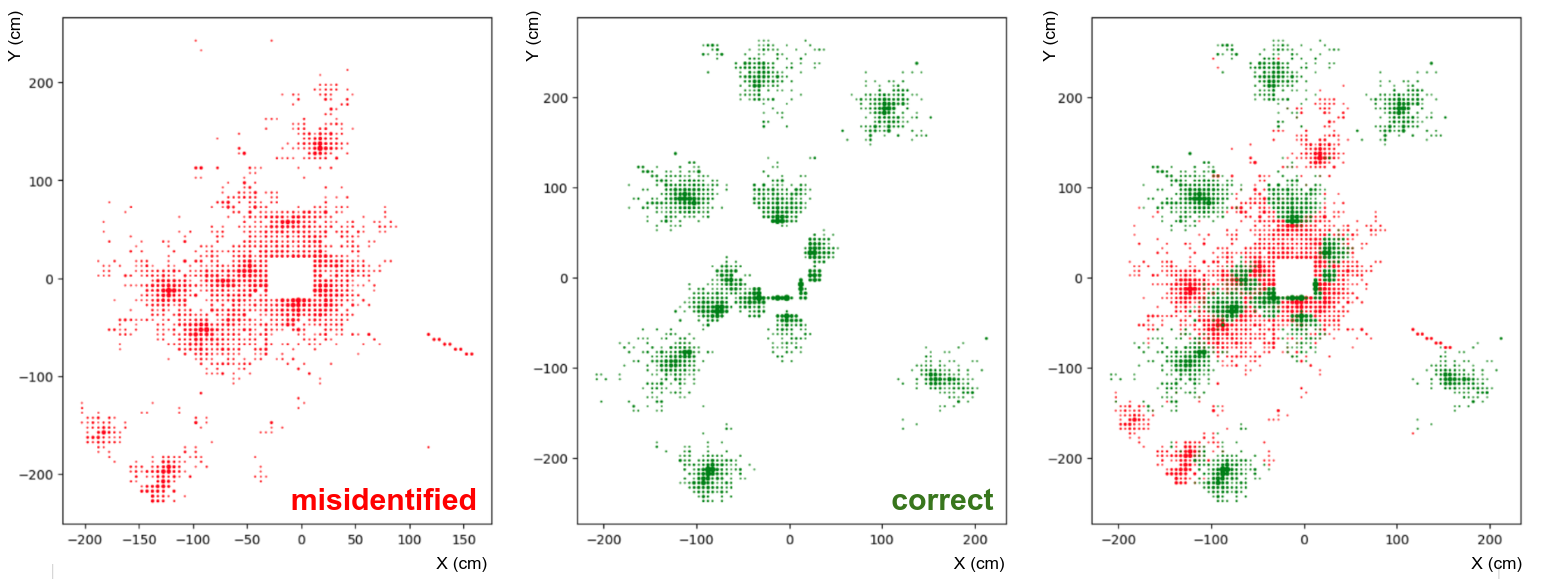}
    \caption{Illustration of the cluster assignment for a specific event with a radius of $R=0.4$ projected on the $xy$ plane. The mislabelled clusters are shown in red, whereas the correctly assigned clusters are shown in green. Figure produced by Group \textit{G}.}
    \label{fig:cluster_assignment_kt_alogithm}
  \end{center}
\end{figure}

The performance of this method as a function of the dimensionless ``jet" radius parameter $R$ for the differing performance metrics such as the completeness, the homogeneity, and the V-score (left) as well as the average energy resolution (right) is shown in \autoref{fig:kT_algorithm_vScore_and_egy_reso_final}. For larger radii, the clusters appear more complete, whereas the homogeneity decreases. The V-score peaks around $R \approx 0.4$, which also yields a near-optimal average energy resolution of the clusters. As a result, $R = 0.4$ was chosen as the resolution parameter for this study. Another parameter that influences the anti-$k_{\mathrm{T}}$ clustering algorithm is the power of the energy used when combining the four momenta of two pseudo-jets. The performance of the clustering algorithm was found to not strongly depend on this parameter. To qualitatively characterize the performance of this method once the parameters have been fixed, one can look at the correctly identified and misidentified hits on an event-by-event basis. This is shown for a single event in \autoref{fig:cluster_assignment_kt_alogithm}. In this event, which is representative of a single event, one can see that the algorithm performs best on the outer edges of the calorimeter where the clusters are well separated. However, near the beam line, where the density of clusters is large and clusters tend to overlap, the algorithm struggles to assign the hits to the correct cluster. The large variation in the density of clusters throughout the phase space makes the application of an algorithm with a fixed resolution parameter, like the anti-$k_{\rm T}$ algorithm, challenging. 

Group \textit{G} ended up as the lowest scoring group, as shown in \autoref{tab:perfComp}.
In order to improve the performance of this non-ML based approach, one could make some small adjustments as to how the algorithm is applied. One option would be to perform the clustering in two stages where firstly the clusters are formed in each z-layer and then combined across layers. Another option would be to cluster layer-by-layer, but incrementally add z layers where the corresponding hits are added to clusters found in the previous layers. Such improvements are left for future work.

\subsection{K-means}

K-means is a popular clustering algorithm that aims to partition data into a set of $k$ clusters. The number of clusters needs to be predetermined, as discussed extensively below. In this algorithm, $k$ cluster centres are initialised at random points within the observation boundaries. The Euclidean distance between the cluster centres and data points are calculated. The data points are then assigned to the cluster centre they are closest to. The central points of the clusters are then calculated. As the cluster centre is updated, the data points need to be reassigned to their closest cluster centre again. This process repeats until the algorithm converges. There is no guarantee that the algorithm will converge on the global optimum, and the algorithm also depends on the position of the initial cluster centres. Since the algorithm is typically quite fast, it is common to repeat the process 
and use the solution where the clusters are on average closest to their cluster centres. 
K-means is a simple and relatively fast algorithm, making it suitable for a wide range of applications. The perhaps biggest drawback of the algorithm is the need to predetermine the number of clusters without any knowledge of the underlying structure within the data. The simplest method is to set a fixed number of clusters that never changes. This was done in the benchmark algorithm, where it was set to 40. This is clearly not an optimal solution, and was chosen so the students could easily beat the benchmark. 

Though many approaches were explored, K-means performed consistently well, so all the final submissions made in this competition apart from submissions \textit{D} and \textit{G} used K-means. They differed mainly by how they determined the number of clusters. 

\begin{figure}[H]
    \centering
    \includegraphics[scale=0.2]{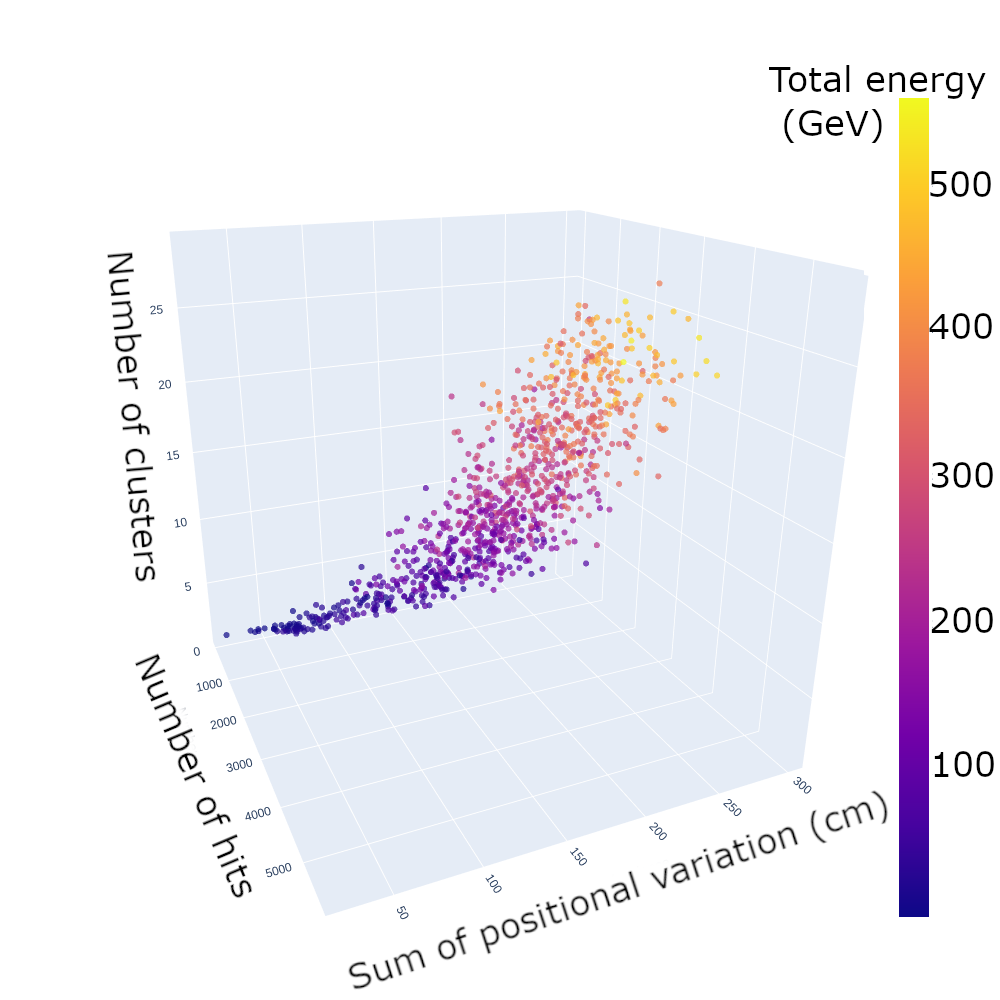}
    \caption{Distribution of the cluster number in a subset of the training data. The number of hits and energy in an event is a good predictor for the number of clusters. Figure produced by Group \textit{E}.} 
    \label{fig:clusterCorrelation}
\end{figure}

One of the key insights that emerged independently from many of the groups is that the optimal number of clusters depends largely on the number of hits in the event. This is shown in \autoref{fig:clusterCorrelation}. Higher energy events are also spread out more in space, meaning they have more clusters. Several of the groups picked up on these patterns. There were many different approaches to finding the optimal cluster number with the most popular methods being regression methods.

\subsubsection{Regression models}
Regression is a generic term for relating one independent variable to one or more dependent variables. The simplest regression model is linear regression, where a linear relationship between the variables are assumed. Some students also explored non-linear regression, motivated by the slight curve in \autoref{fig:clusterCorrelation}. Group \textit{C} employed linear regression using the number of hits and total energy of the event. As can be seen by the scores in \autoref{tab:perfComp}, this performed fairly well. Group \textit{E} also included the positional variation in each dimension of the event in their regression. Perhaps surprisingly, this did not outperform Group \textit{C's} simpler solution. The variation in each coordinate is quite related to each other, and they are also correlated to the total energy in the event. Having correlated variables is known to potentially decrease accuracy in linear models \cite{kuhn2013applied}. The highest performing solution was \textit{A}'s quadratic regression using the number of hits only. This is quite motivated by \autoref{fig:clusterCorrelation}.

\subsubsection{Neural Networks}
Neural networks, or neural nets, combine linear regression with non-linear functions into a connected network of nodes, somewhat inspired by neurons in the brain \cite{Goodfellow-et-al-2016}. Because of the non-linearity and number of parameters, neural nets are suitable for a wide range of tasks, and are the go-to approach for any particularly complex problem, like image classification or language processing. 
Group \textit{E} tried to predict the number of clusters using a neural net of three to five layers. Their idea was that this could capture special cases that deviate slightly from the central trend in \autoref{fig:clusterCorrelation}. This approach did not outperform a linear regression, and the neural net had a larger variance in the number of predicted clusters than the regression models. This is shown in \autoref{fig:linear-nn}, where the Figure was made by Group \textit{F}, who did something very similar. The neural net produces results similar to a linear regression, but is arguably slightly overfit. Neural nets are great for complex data, but this data, especially when expressed with only one explanatory variable is not very complex, and does probably not warrant a neural net. 

\subsubsection{DBSCAN}\label{dbscan}
A very different approach to finding the right number of clusters was the DBScan (Density-Based Spatial Clustering of Applications with Noise) algorithm \cite{dbScan}. This is a clustering algorithm that does not require a pre-set number of clusters. Instead, data is split into clusters based on the density of the region. The density is defined by counting neighbours that are within a pre-defined distance from each other. The groups that explored DBSCAN all concluded that it did not work as well as K-means as a standalone algorithm. A potential reason for this could be that the distance between detector hits in a cluster will be smaller closer to the beamline, and spread out further in the detector. Most of the groups instead tried to use DBSCAN as a way to determine the number of clusters to be used in K-Means. Most concluded that not only did this take quite long, it also didn't outperform a simple linear regression.  

\subsubsection{Ensemble Methods}
\begin{figure}[ht!]
    \centering
    \includegraphics[width=0.5\linewidth]{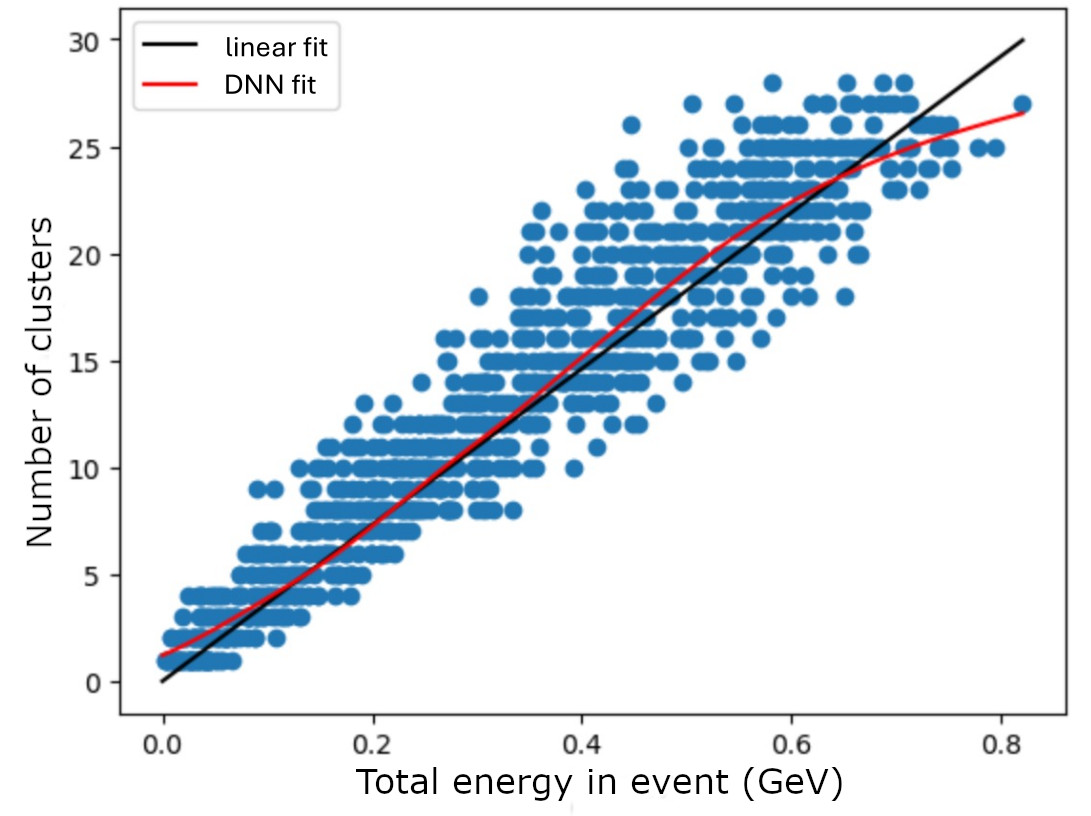}
    \caption{Number of clusters versus the total energy in an event including fits determined by linear regression (black) and a neural network (red). Figure produced by Group \textit{A}.}
    \label{fig:linear-nn}
\end{figure}

Ensemble methods aim to combine a set of weak learners into a stronger learner~\cite{10.1007/3-540-45014-9_1}. This is typically done with either majority voting or boosting. In majority voting, the output of the models are combined into a final prediction. This if often done by averaging the output, or choosing the most frequent output. In boosting, the errors of one model are emphasised for the training of the next model, thereby gradually improving the models. Group \textit{F} combined a neural net with a linear fit to determine the number of clusters. The fit of the neural net and linear regression to the number of clusters is shown in \autoref{fig:linear-nn}. From the Figure, it is clear to see that in most cases, using this ensemble method to determine the number of clusters will have very little effect, as the models are largely similar. The neural net differs slightly at very low and high energy, and is arguably slightly overfit. 

Group \textit{E} used ensemble learning in a somewhat untraditional way. They observed that their model seemed to perform better with some features for certain events, and another set of features for other events. K-means uses a measure of how spread out a cluster is called inertia \cite{kmeansInertia}. Group \textit{E} performed clustering with two sets of features, and chose the one with the lowest inertia. They saw that this improved their V-score, suggesting that inertia is related to the V-score. 

\subsubsection{Pre- and postprocessing}
Finding the optimal number of clusters for K-means algorithms took up much of the development time during the Power Week. There were also some major advancements during pre- and postprocessing of the data. The most notable development was using spherical instead of Cartesian coordinates. Once this was suggested, most groups adapted it and saw an increase in their classification scores. All groups apart from Group \textit{G} and Group \textit{E} used this in their final submissions. One can perhaps view the propagating particles as a wavefront, making it more suitable for a spherical description. K-means uses a Euclidean distance to calculate the distance between cluster centres and the data points. This is not as suitable for spherical data; points that are close on the surface of a sphere are not necessarily close by Euclidean distance. It would have been interesting to try a different distance metric, such as the great-circle distance, to see if this would improve the results. Some groups also tried to normalise the data, but K-means in \texttt{scikit-learn}~\cite{scikit-learn} normalises the data in the back-end, so this had no effect.

\subsection{Autoencoder}
\begin{figure}[ht]
    \centering
    \includegraphics[width=0.49\linewidth]{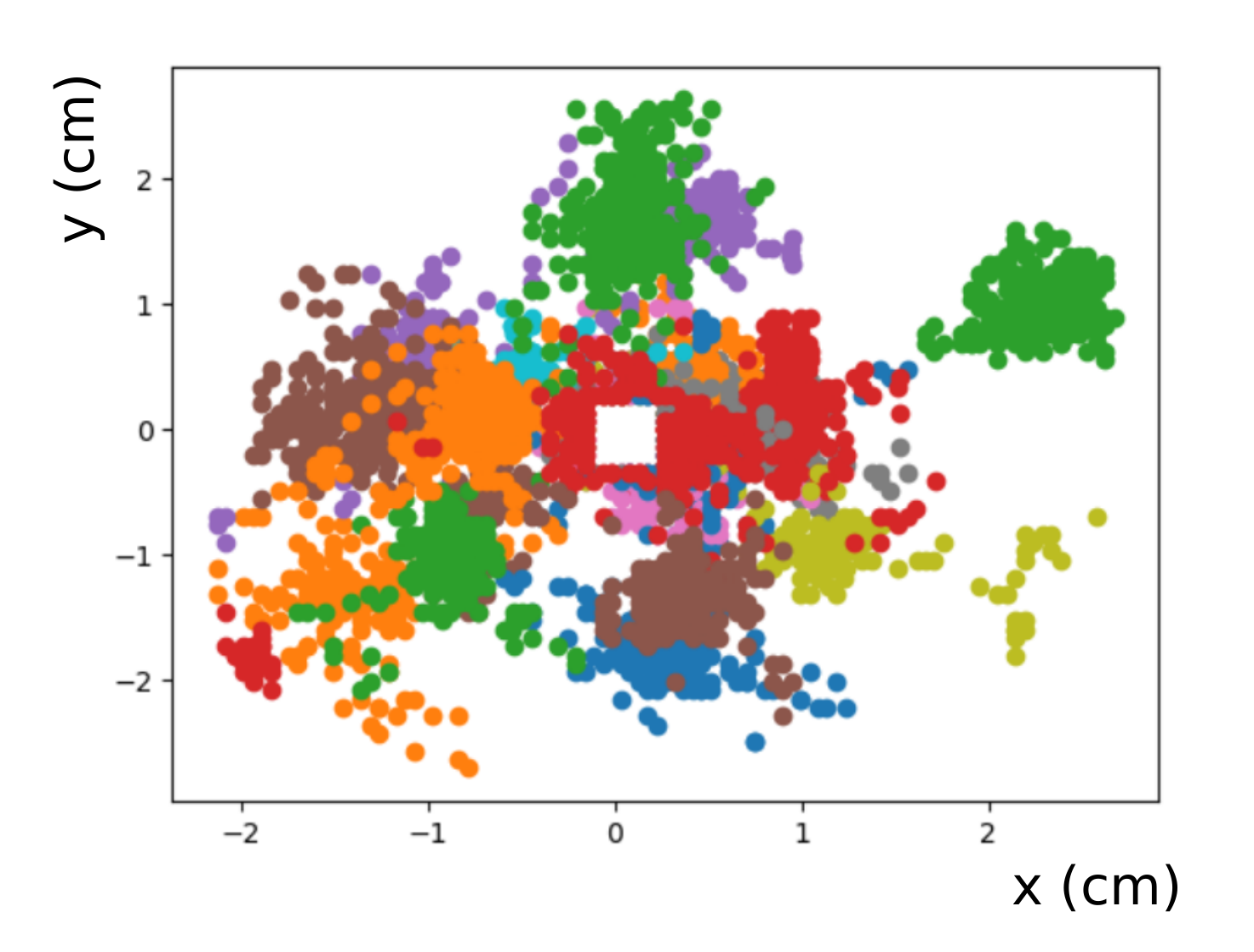}
    \includegraphics[width=0.475\linewidth]{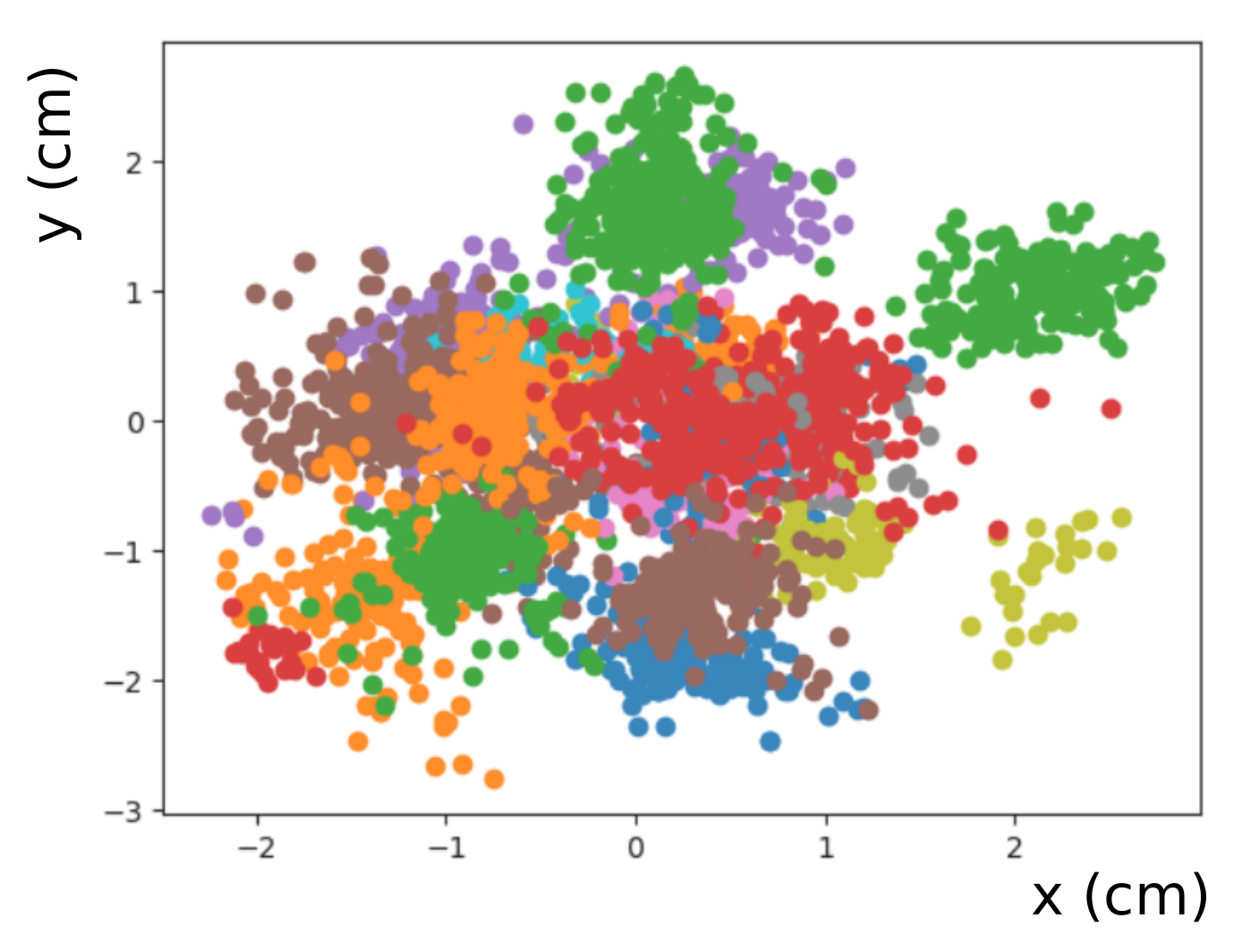}
    \caption{(Left) Hits and their corresponding cluster identifications for an example event passed as input to the autoencoder. (Right) Hits and their corresponding cluster identifications for a reconstructed event from the autoencoder. Figure produced by Group \textit{G}.}
    \label{fig:autoencoder}
\end{figure}

Autoencoders typically consist of two neural nets that work together to learn to compress and reconstruct a space~\cite{Goodfellow-et-al-2016}. The first neural net learns an embedding of the input features by passing it through the neural net. The second neural net acts as a decoder to reconstruct the embedded space back to the physical space. The two neural nets share a loss function that measures the difference between the input data and the decoded space. This can be very effective for compressing data, finding important features, or identifying  anomalies in data. Groups \textit{G} and \textit{E} attempted to use autoencoders during this project, with the motivation that the autoencoders could learn to represent the clusters in a space where the clusters are better separated. Both found that it did not outperform the K-means algorithm. This is shown in \autoref{fig:autoencoder}, where one can see by eye that the autoencoder does not visibly improve the cluster separation.

\subsection{Graph-based model}
Only one solution used a graph-based approach, namely Group~\textit{D}. The algorithm is motivated by the success of graph-based segmentation and reconstruction techniques in other particle physics applications, such as \cite{ju_graph_2020,Caillou:2815578, ATL-PHYS-PUB-2022-027, duperrin2023flavour, Abbasi_2022, Bhattacharya:2803236} (full reviews given in \cite{thais2022graph, Shlomi_2021}). In particular, it applies a graph structure to the problem, such that any edge classifier, whether a simple multi-layer perceptron (MLP), GNN or transformer, can be used to prune spurious connections between a set of high-energy seeds and the rest of the energy deposits.

The algorithm works in two stages: a) seed-to-seed classification, then b) seed-to-nonseed segmentation. The first stage begins by defining for each event of $N_{h}$ hits, some $N_{s}$ number of seeds, taken to be the $N_{s}=40$ highest-energy hits in the prototype implementation. A fully-connected graph is constructed between these $40$ seeds, with $N_{s}^{2}=1600$ edges to be classified. An edge connecting two seeds from the same shower source is given a ground truth label of 1, otherwise it is given a label of 0. A simple MLP is trained on the 1600 edges across 20k events, with input features given by the two seeds belonging to the edge.

During inference, edges are scored by the MLP, with edges below some threshold (taken to be $0.6$) discarded, and edges above the threshold retained as $e^{seed}_{seed}$.

Next, the remaining $N_{h} - N_{s}$ hits are each connected with an edge to their nearest seed. This can be achieved by applying a K-nearest neighbour algorithm with K=1, with the database and query set given by the $N_s$ seeds and the $N_{h} - N_{s}$ nonseeds, respectively. This returns a set of $N_{h} - N_{s}$ edges $e^{seed}_{nonseed}$. To obtain unique cluster labels to each hit, we apply a connected components algorithm to the full graph.

The connected components algorithm performs a depth-first search from each node to all other nodes that can be reached by an edge, assigning the set a common label. Looping this procedure produces component labels for all nodes. A high-performing algorithm exists~\cite{Pearce2005AnIA}, such as that used here from the \texttt{SciPy} library~\cite{2020SciPy-NMeth}.

\begin{figure}[ht!]
    \centering
    \includegraphics[width=0.4\linewidth]{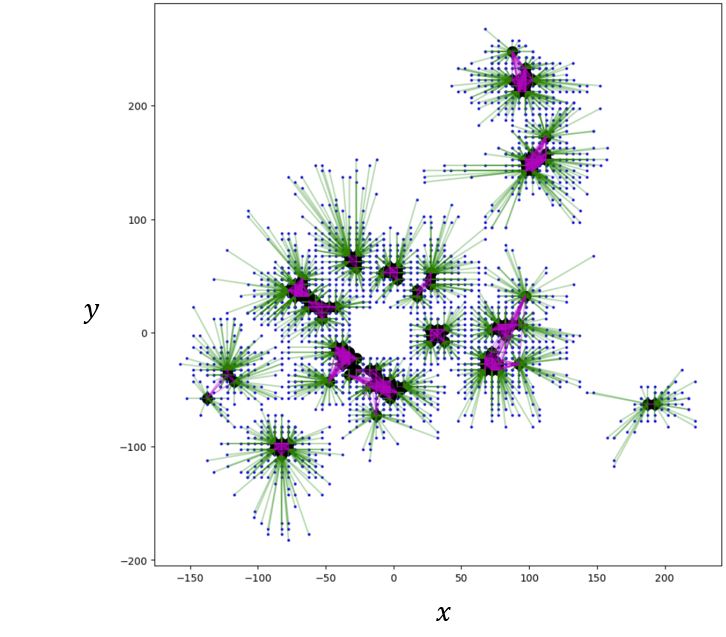}
    \includegraphics[width=0.4\linewidth]{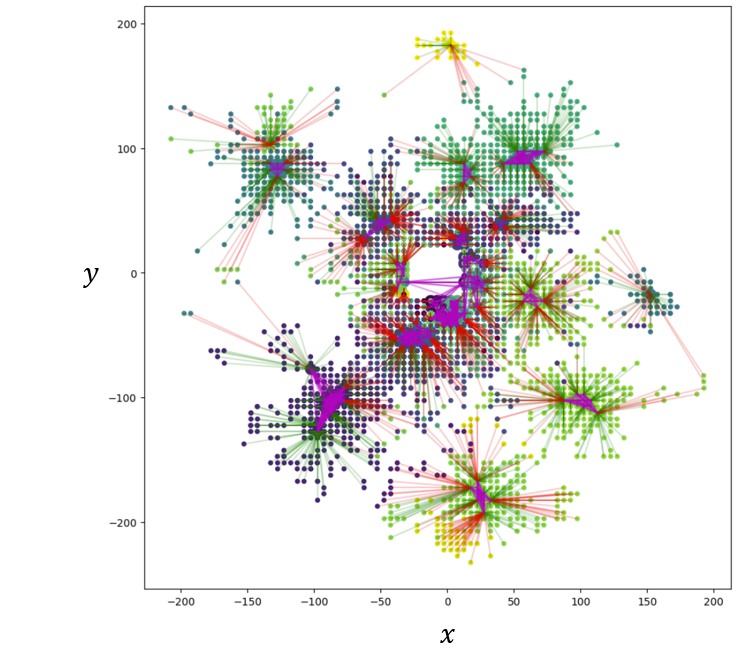}
    \caption{Example graphs constructed on showering events. (Left) The outcome of the first stage, where seeds (large black points) are connected to each other if their shared edge passes a classification prediction by an MLP. (Right) The outcome of the second stage, where nonseeds (small points coloured by particle ID) are connected to their nearest neighbouring seed in 3D space. Figure produced by Group \it{D}.}
    \label{fig:graphcluster}
\end{figure}

The implementation of the algorithm described above was able to achieve a middle-of-the-pack V-score of 0.96570. The output on an example event is given in \autoref{fig:graphcluster}. We may inspect the performance specifically of the edge classifier by defining an edge-wise efficiency (ratio of true positive edges to true edges) and purity (ratio of true positive edges to positive edges). The core component of this model, the edge-classifier MLP, achieved a best efficiency of 0.956, purity of 0.956, and an area under the receiver operator characteristic curve (AUC) of 0.998.

Submission \textit{D} as presented here exhibited competitive performance with unsupervised clustering, but did not achieve state-of-the-art results. There are, however, a variety of improvements that could be made to the model and training process that should allow the model to produce better results. The model itself is very simple, composed only of an MLP followed by the K=1 KNN seed-to-nonseed connection and a brute-force connected components clustering. In the first instance, the MLP could be trivially replaced by a GNN or Transformer, giving the classifier more local or global context. The clustering performance of later steps is highly dependent on this edge classifier, and even incremental improvements to this stage should meaningfully improve downstream performance. Finally, due to the time constraints of the competition, only 20,000 events were trained on, with overfitting observed. One may expect better performance with a larger training dataset.

\section{Discussion and Recommendations}\label{sec:discussion}
A comparison of the weighted V-score for each of the submissions can be found in \autoref{tab:perfComp}, giving a sense of the relative performance of each approach. However, as explained previously, there are many possible methods that can be used to compare the performance of clustering algorithms. Such methods as well as general recommendations will be discussed in the following subsections.

\begin{table}[ht!]
    \centering
    \begin{tabular}{c|c}
        \textbf{Submission} & \textbf{Weighted V-Score}\\ \hline
         Benchmark & 0.95207\\ 
         (A) & 0.96692\\
         (B) & 0.96688\\
         (C)& 0.96668\\
         (D) & 0.96570\\
         (E) & 0.96453\\
         (F) & 0.96054\\
         (G) & 0.95395\\
    \end{tabular}
    \caption{Performance of the various submissions by weighted V-Score.}
    \label{tab:perfComp}
\end{table}

\subsection{Comparison of the solutions}
Most of the solutions presented in this paper use K-means with some method to select the value of $k$, and it seems reasonable that this would lead to somewhat similar solutions. We will now compare the solutions with metrics beyond the V-score, including the benchmark. We did not have access to the solution from Group \textit{B}, so this is not included in these plots. 

To compare solutions, we need to decide on a way to match different clusters since each group's cluster IDs are arbitrary. There are perhaps two main approaches to this. One could match the truth and predicted clusters by either majority or energy. In the former, cluster A would be matched to truth cluster B if most of the hits in A belong to truth cluster B. In the latter, we ignore the majority, and instead consider the most energetic hit. If the most energetic hit in A belong to truth cluster B, then we say all the hits in A were predicted to be in B. Both approaches can be valid, and are in this case largely equivalent. 

\begin{figure}[ht]
        \centering
        \includegraphics[width=0.4\linewidth]{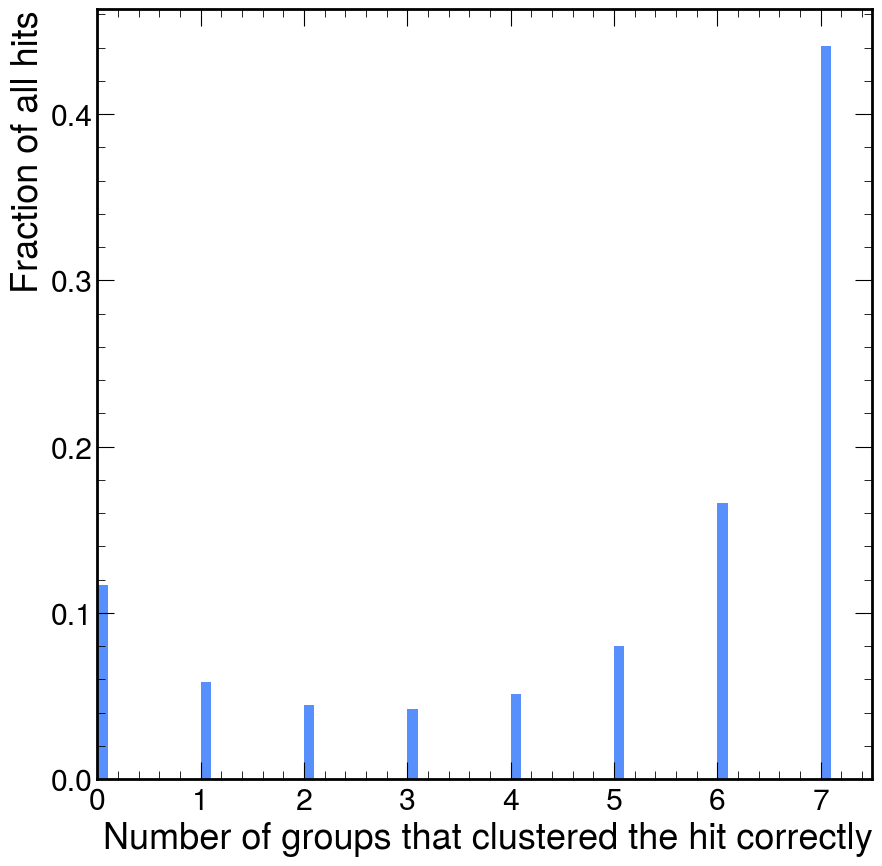}
        \includegraphics[width=0.4\linewidth]{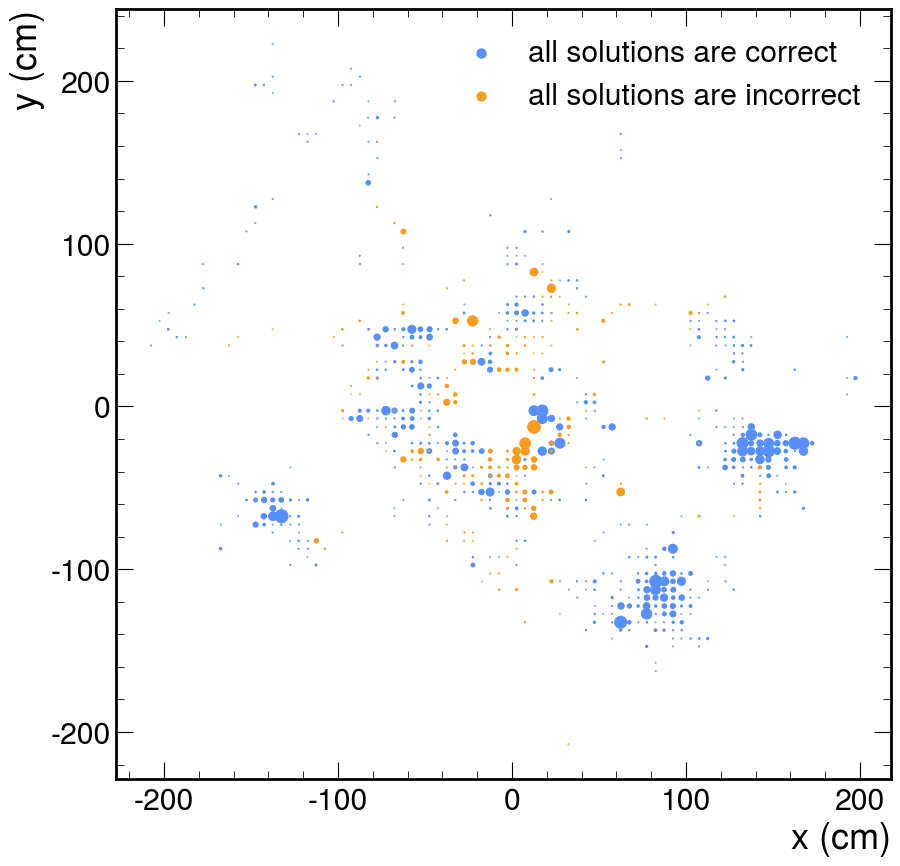}
=        \caption{(Left) The ratio of hits for which n solutions got the
correct solution aggregated over 1000 events
from the test data. (Right) The distribution of hits where all the solutions
agree and are correct, and where they are all
incorrect for a randomly chosen event.}
        \label{fig:spread_misclassification}
\end{figure}

Here, we chose to compare the solutions based on the majority. One thousand events from the test data were used for these comparisons. We can see that the fraction of times that all the solutions agree \textit{and} are correct is almost 50\% of the time, as shown in \autoref{fig:spread_misclassification}.

There are certain situations that all the approaches struggle with. This is shown in \autoref{fig:spread_misclassification}. This Figure is based on a randomly selected event, but is representative of the dataset as a whole. 
In most cases, the algorithms struggle when the hits are close to the beamline. This holds true even if we consider this relative to the density of the region. In the test sample of 10,000 events, 97\% of the cases where all the solutions were wrong, the hits were within $\pm$ 80 cm from the origin in the $x$ and in the $y$ direction. Comparing \autoref{fig:spread_misclassification}, we can see that even though most of the solutions use K-means, the small variations can lead to meaningful differences. From \autoref{fig:spread_misclassification}, we see that at least a 10\% classification accuracy is lost for all the solutions because of the difficultly of classifying hits near the beamline. The largest performance gain would therefore be to find a way to better classify these hits. 

\begin{figure}
    \centering
    \includegraphics[width=0.5\linewidth]{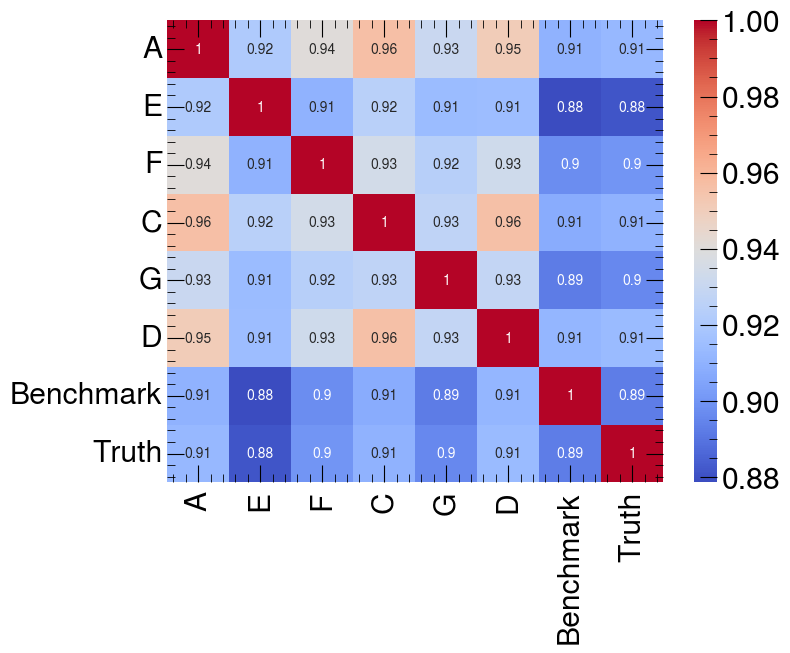}
    \caption{Heatmap of the normalised mutual information score.}
    \label{fig:solution_heatmap}
\end{figure}

Since all the submissions apart from two use K-means, we would like to know whether the graph-based or Anti-{$k_t$} approach pick up on fundamentally different patterns in the data. We can measure this with the Normalised Mutual Information (NMI) score \cite{vinh2009information}. This measures the similarity of two clustering solutions by considering how much information one cluster contains about the other, and then normalizing it so that the cluster size doesn't matter. \autoref{fig:solution_heatmap} shows a comparison between the solutions available. Again note that submission \textit{B} was not available when comparing the solutions. By comparing this Figure to the results in \autoref{tab:perfComp}, we see that the top scoring solutions - \textit{A}, \textit{C} and \textit{D} share a very high NMI score. This illustrates that the top scoring solutions all found similar patterns in the data. There is no evidence that the graph-based or Anti-{$k_t$} solutions, i.e. submission \textit{D} and \textit{G} respectively, found different patterns than the K-means approaches. 
We would also expect that the lower scoring solutions have a lower NMI score when compared to the truth solution. The lowest score is solution \textit{E}, but the lowest scoring solution in the Kaggle competition was \textit{G}. This goes to show that the choice of scoring metric leaves ambiguity, as discussed in the following section. 

\subsection{Scoring Metric}
In general, providing a single figure of merit for the reconstruction of a physical phenomenon is an underconstrained problem. The downstream use and the environment (i.e. offline or online) of the reconstruction will highly influence which behavior to encourage in an algorithm. Clustering in particular is a subtle task to evaluate, as evidenced by the nine independent sets of metrics provided by \texttt{scikit-learn}~\cite{scikit-learn}. However, choosing to use a single metric, and one that was not motivated by a particular downstream task, turned out to be essential for students to gain intuition about the problem and quickly iterate on solutions. The process of \textit{designing} this metric as one guided by the physical phenomenon (e.g. evaluating cluster assignment weighted by energy) was also very useful for the organizers, aiding in iterating on the problem statement and the dataset features provided.

As a calorimeter clustering challenge that was agnostic to physics analysis application, some version of the weighted V-score is indeed ideal. Refinements would be motivated by how high-scoring even simple algorithms performed. This is both a reflection of the dataset being relatively ``easy", and the behavior of the V-score to approach 1.0 as the number of clusters increased. As such, the scoring function provided in the challenge, which merged all events before scoring, gave an artificially high score. For example, a V-score across 5000 event predictions of one algorithm gave 0.965, while taking the score across 100 events at a time gave an average V-score of 0.433. This behaviour should be corrected in future challenges.

Translating the algorithms implemented in this challenge to realistic physics applications will of course require a more nuanced evaluation. Inference time, while very important for groups' ability to iterate quickly, was not included in the figure of merit in the challenge. Multiplying the V-score by some function of time (e.g. $e^{-t}$) may be a useful approach. Benchmarking the various algorithms with both the physics-agnostic V-score and some downstream task-focused metrics, for example $|E_\text{reconstructed} - E_\text{reconstructable}|/E_\text{true}$, would be sensible in order to establish a correspondence.

\subsection{Improvements to Format}
In order to aid future ``Power Weeks" targeted at ML development, whether they be open workshops aimed at students acquiring ML knowledge (like this one) or a ``workfest" amongst experiment experts, there are a few improvements to the format used here that the authors would recommend. Firstly, we would recommend that a wider variety of ML-based examples be prepared before the week begins such that the participants can focus on studying the performance of different algorithms versus spending the time to implement them. For example, due to the short time available to optimize the methods to the problem, most participants utilised the K-means algorithm first made available to participants as an example. However, with a more diverse pool of examples, one would also expect a more diverse pool of solutions. Continuing in this vein, the authors would also recommend that there be some changes to the format utilised here in order to promote both \textit{exploration} and \textit{optimization}. For example, one could consider a more spread out format (i.e. longer than one week) with two intense working periods. The first of these two periods would occur for 2-3 days at the beginning of the event and consist of an intense data exploration and review of some examples of ML methods. Then 2-3 weeks with low intensity lectures and individual group meetings. Finally there would be an additional 3 day period at the conclusion of the event for last minute optimization and in-person discussions. In addition, providing data samples with a smaller number of events would greatly reduce processing time and allow for quicker developments~\footnote{Such an improvement is particularly useful in a ``Power Week" style format as employed here.}. Overall these improvements would allow for the overall quality of the solutions to be higher and for students to learn more about how ML applications work in practice. 

\subsection{Detector Design}
This Power Week provided a unique opportunity where a feedback loop can be formed between detector design and algorithmic performance. Therefore, the detector optimization can be formed with ease of clustering in mind. For example, no matter the implementation details of the approach, clustering particles around the beamline is an area where algorithms fail to properly identify true clusters. Though this will always be a difficulty, as a majority of particles are produced incredibly close to the beamline, this exploration shows that optimizing detector performance and granularity would be useful in this region. In addition, one can see that from the clustering perspective, fine segmentation in $z$ is not critical. These considerations were presented to the LFHCAL working group of the ePIC collaboration in August of 2023~\footnote{\href{https://indico.bnl.gov/event/19752/contributions/79741/attachments/49155/83760/ePIC_CaloWG_Meeting.pdf}{https://indico.bnl.gov/event/19752/contributions/79741/attachments/49155/83760/ePIC\_CaloWG\_Meeting.pdf}}. Though there are many aspects to consider when performing an optimization of detector design, this challenge was able to provide some early feedback, before detector design was finalised, as to which attributes of detector design was most important for clustering.

\section{Summary and Conclusions}\label{sec:summary}
The summary of scores of the above-mentioned approaches can be found in Table \ref{tab:perfComp}. Note that all teams were able to improve in performance over the benchmark. There were a variety of approaches to this problem that ranged from ``traditional approaches" (anti-$k_{\rm T}$), modified $K$-Means, and supervised learning. This variety is a testament to the idea that crowd-sourced development is an effective tool to generate creative solutions to long-standing problems. 

Overall, this program allowed participants to learn about machine learning while also actively contributing to an unsolved problem in high energy physics. The details presented in this work aim to not only record the feedback on similar ``Power Week" structures, but also summarize the insights into clustering problems for future forward calorimeters at the EIC.

\begin{sidewaystable}[ht!]
\centering
\begin{tabular}{|p{2cm}|p{3.1cm}|p{3cm}|p{2cm}|p{1.5cm}|p{2cm}|p{1.5cm}|p{2cm}|p{2cm}|p{1cm}|}
\hline
\textbf{Name} & \textbf{Algorithm, $N$-finding Method} & \textbf{Inputs to find $N$} & \textbf{Coordinate System} & \textbf{Kmeans Inputs} & \textbf{Init.} & \textbf{Sample Weight} & \textbf{Ensembling} & \textbf{Post-processing} & \textbf{$V_{score}$} \\ \hline
HelloWorld (A) & K-Means, Quadratic Regression & Number of active cells & Spherical & $\phi, \theta$ & k-means++ & - & Inertia scan around prediction & - & 0.96692 \\ \hline
The Winning Team (B) & K-means, Linear regression & Number of hits & Spherical & $\phi, \theta$ & k-means++ & - & - & - & 0.96688 \\ \hline
ClusterDuck (C) & K-means, Linear Regression & Total number of hits, total deposited energy & Spherical & $\phi, \theta$ & - & Energy (E) & - & Basis transformation to spherical coordinates, energy weighting & 0.96668 \\ \hline
JMeans (D) & Graph neural net & - & Spherical & - & - & - & Inertia scan around prediction & - & 0.96570 \\ \hline
Genuine Stupididy (E) & K-means, Regression (OLS, GLM) \& Neural Network & Number of hits, energy, summed std. deviation & Cartesian & $x, y, z_i$ & k-means++ & Energy (E) & Multiple feature combinations & K-Means bisection for overlaps & 0.96453 \\ \hline
Undergrad Vision (F) & K-means, Ensemble (Linear Regression \& Neural Network) & Total energy, num. hits & Spherical & $\phi, \theta$ & k-means++ & - & Neural Network weighted twice & - & 0.96054 \\ \hline
OldRabbits (G) & Anti-$k_{t}$ & - & Cartesian & $x, y, z_i$ & jet radius = 0.4 & - & - & - & 0.95395 \\ \hline
Benchmark & None - constant N & N/A & Cartesian & $x, y, z_i$ & k-means++ & - & None & - & 0.95207 \\ \hline

\end{tabular}
\caption{Comparison of Different Algorithms}
\label{tab:kmeans_algo_comparison}
\end{sidewaystable}

\subsection{Feedback}
Following the course, participants were asked to fill out a survey asking open-ended questions on the learning objectives, learning environment, course content and style, and other relevant topics.  One common theme in the feedback forms was that participants wished there to be more complicated ML architectures used in the example code. The K-means example provided was meant to provide a simple entry point for participants. However, an unintended consequence of this working well was that participants did not get to learn as much about more complex ML techniques, such as autoencoders or CNNs. In addition, the environment of the course was quite remote, which allowed for an immersive experience, but did not provide as accessible access to wifi for example, which made some aspects of the course more difficult. Overall, the response from participants was generally positive and demonstrated reported a deeper understanding on machine learning than before taking the course.

\section{Acknowledgments}
The authors would like to thank the Helmholtz Graduate School for Hadron and Ion Research and its partners for their generous support of this competition. We would also like to thank the ePIC collaboration, specifically Nicolas Schmidt and Friederike Bock (ORNL), for providing the simulations used in this challenge as well as advice throughout. 
\newpage
\bibliography{ref}
\end{document}